\documentclass[a4paper]{jpconf}
\usepackage{graphicx}
\begin{document}
\title{An evolving hot spot orbiting around Sgr~A*}

\author{M~Zamaninasab$^{1,2}$, A~Eckart$^{1,2}$, L~Meyer$^{3}$, R~Sch\"{o}del$^{4}$,  M~Dovciak$^{5}$, V~Karas$^{5}$, D~Kunneriath$^{1,2}$, G~Witzel$^{1}$,  R~Gie{\ss}\"{u}bel$^{1}$, 
  S~K\"{o}nig$^{1}$,
  C~Straubmeier$^{1}$ and A~Zensus$^{2,1}$}

\address{$^{1}$ University of Cologne, Z\"{u}lpicher Str. 77, D-50937 Cologne, Germany\\
$^{2}$ Max-Planck-Institut f\"{u}r Radioastronomie, Auf dem
H\"{u}gel 69,
53121 Bonn, Germany\\
$^{3}$ Department of Physics and Astronomy, University of
California, Los Angeles, CA~90095-1547, USA\\
$^{4}$ Instituto de Astrof\'{\i}sica de Andaluc\'{\i}a, Camino Bajo de Hu\'{e}tor 50, 18008 Granada, Spain\\
$^{5}$  Astronomical Institute, Academy of Sciences,
         Bo\v{c}n\'{i} II, CZ-14131 Prague, Czech Republic\\
} \ead{zamani@ph1.uni-koeln.de}

\begin{abstract}
Here we report on recent near-infrared observations of the Sgr~A*
counterpart associated with the super-massive $\sim4\times10^6M_{\odot}$
black hole at the Galactic Center. We find that the May 2007 flare
shows the highest sub-flare contrast observed until now, as well as
evidence for  variations in the profile of consecutive
sub-flares. We modeled the flare profile variations according to
the elongation and change of the shape of a spot due to differential rotation within the accretion disk.
\end{abstract}

\section{Introduction}
Over the last decades, evidence has been accumulated that most
quiescent galaxies harbor a massive black hole (MBH) at their center.
Especially in case of the center of our Galaxy, progress has been made
through the investigation of stellar dynamics (Eckart \& Genzel 1996,
Genzel et al. 1997, 2000, Ghez et al. 1998, Eckart et al. 2002,
Sch\"{o}del et al. 2002, 2003, Eisenhauer 2005) which has revealed the
presence of a super-massive $\sim 4\times 10^6 M_{\odot}$ black hole
at the Galactic Center. Its position coincides with that of the
compact radio source Sgr~A*. At a distance of only $\sim 8$ kpc
(Eisenhauer et al. 2005) the Galactic Center is the closest galactic
nucleus. Sgr~A* is a source of variable emission in the X-ray and
near-infrared wavelength regimes (Baganoff et al. 2001, Eckart et al.
2004, 2006a, Genzel et al. 2003 and Yusef-Zadeh et al. 2006). The
NIR/X-ray variability is probably also linked to the variability at
radio through sub-millimeter wavelengths, showing that variations
occur on time scales from hours to years.

\begin{figure}[t]
\begin{center}
\begin{minipage}{4cm}
\includegraphics[width=4cm]{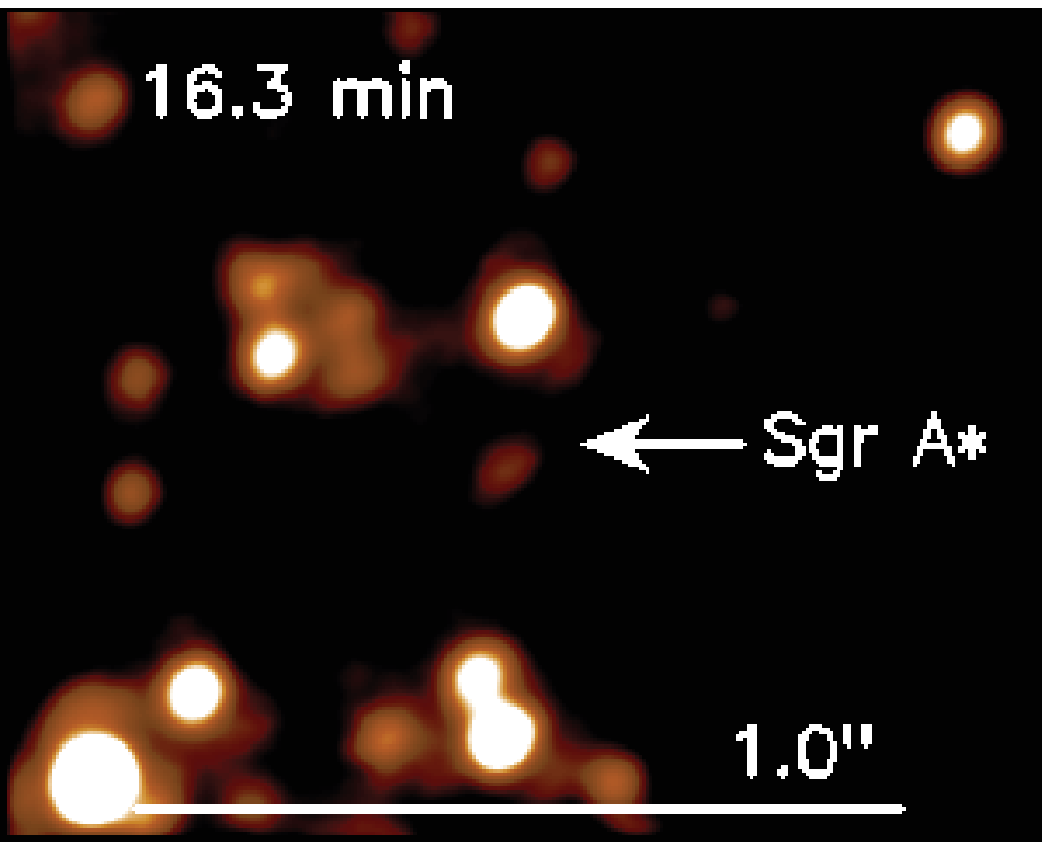}
\end{minipage}\hspace{0.5pc}%
\begin{minipage}{4cm}
\includegraphics[width=4cm]{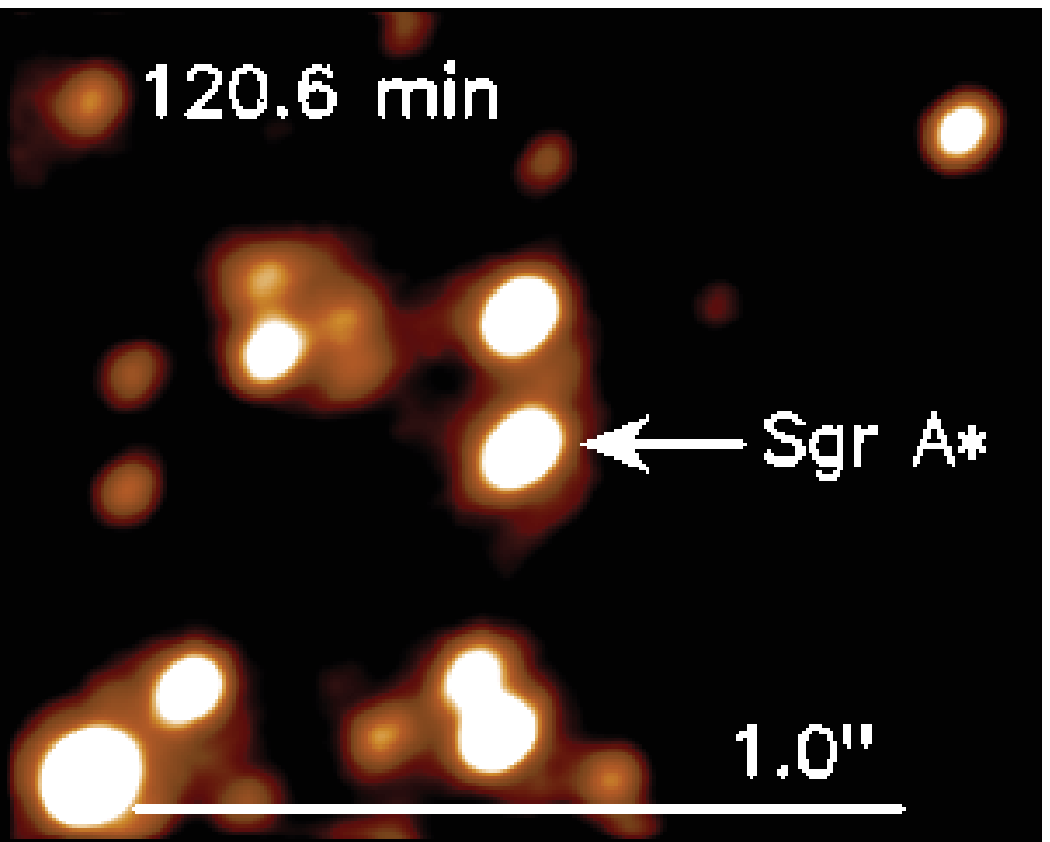}
\end{minipage}\hspace{0.5pc}
\begin{minipage}{4cm}
\includegraphics[width=4cm]{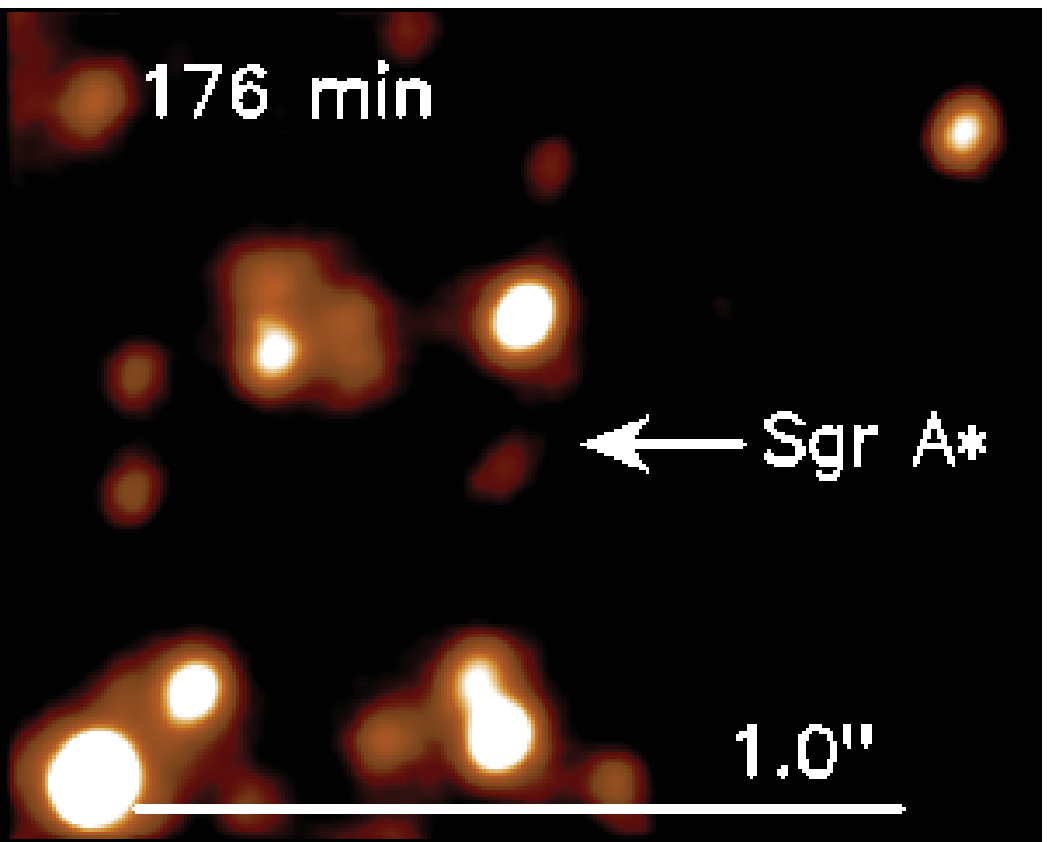}
\end{minipage}
\caption{Sgr A* as it was observed in  NIR $K$-band on
15 May 2007 between 05:29:00 - 09:42:00 ~(UT time) .}\label{moha-Fig0-1}
\end{center}
\end{figure}

Recent NIR and X-ray observations have revealed the non-thermal nature
of the high frequency radiation from Sgr~A*. The source is visible in
NIR regime only during its flaring state
(Fig.~\ref{moha-Fig0-1}). Some NIR flares have been found to be highly
polarized. Most flares normally have X-ray counterparts, which
strongly suggests synchrotron or synchrotron self Compton (SSC)
processes as the radiation mechanism. NIR spectroscopic observations
have shown that the observed spectra of NIR flares can be fitted well
with a power-law, $F_\nu\propto\nu^{-\alpha}$ (where $F_\nu$, $\nu$
and $\alpha$ are the flux density, frequency and spectral index
respectively). Although all observations are in agreement with the
fact that NIR flares show a soft spectrum ($\alpha>0$), the spectral
index value is still not well determined (Eckart et al. 2004,
Gillessen et al.  2006a, Hornstein et al. 2007).

\begin{figure}[t]
\begin{center}
\hspace{-5pc}
\begin{minipage}{4cm}
\hspace{-1.3cm}\includegraphics[width=6cm,angle=0]{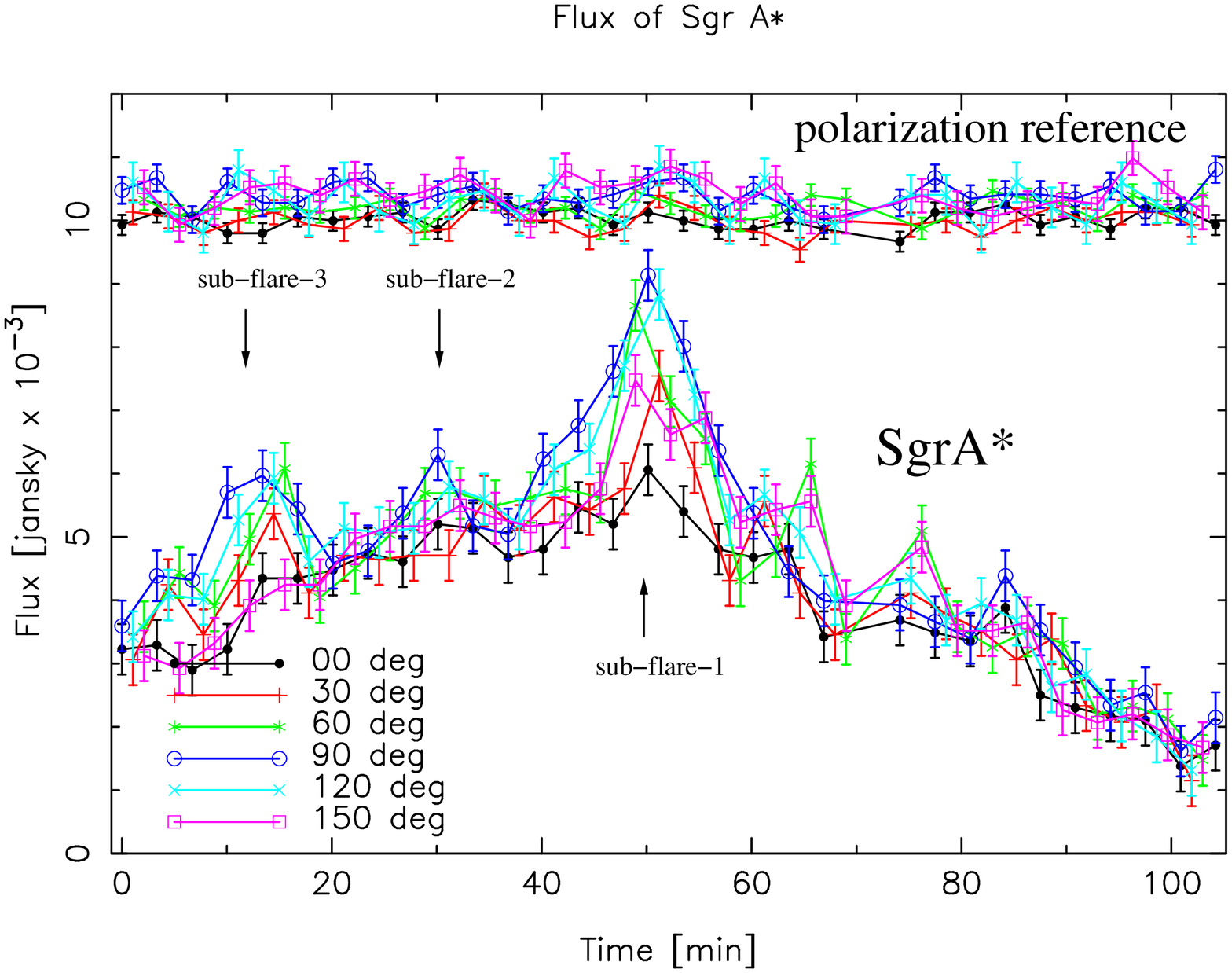}
\end{minipage}\hspace{6pc}%
\begin{minipage}{4cm}
\includegraphics[width=4.7cm,angle=270]{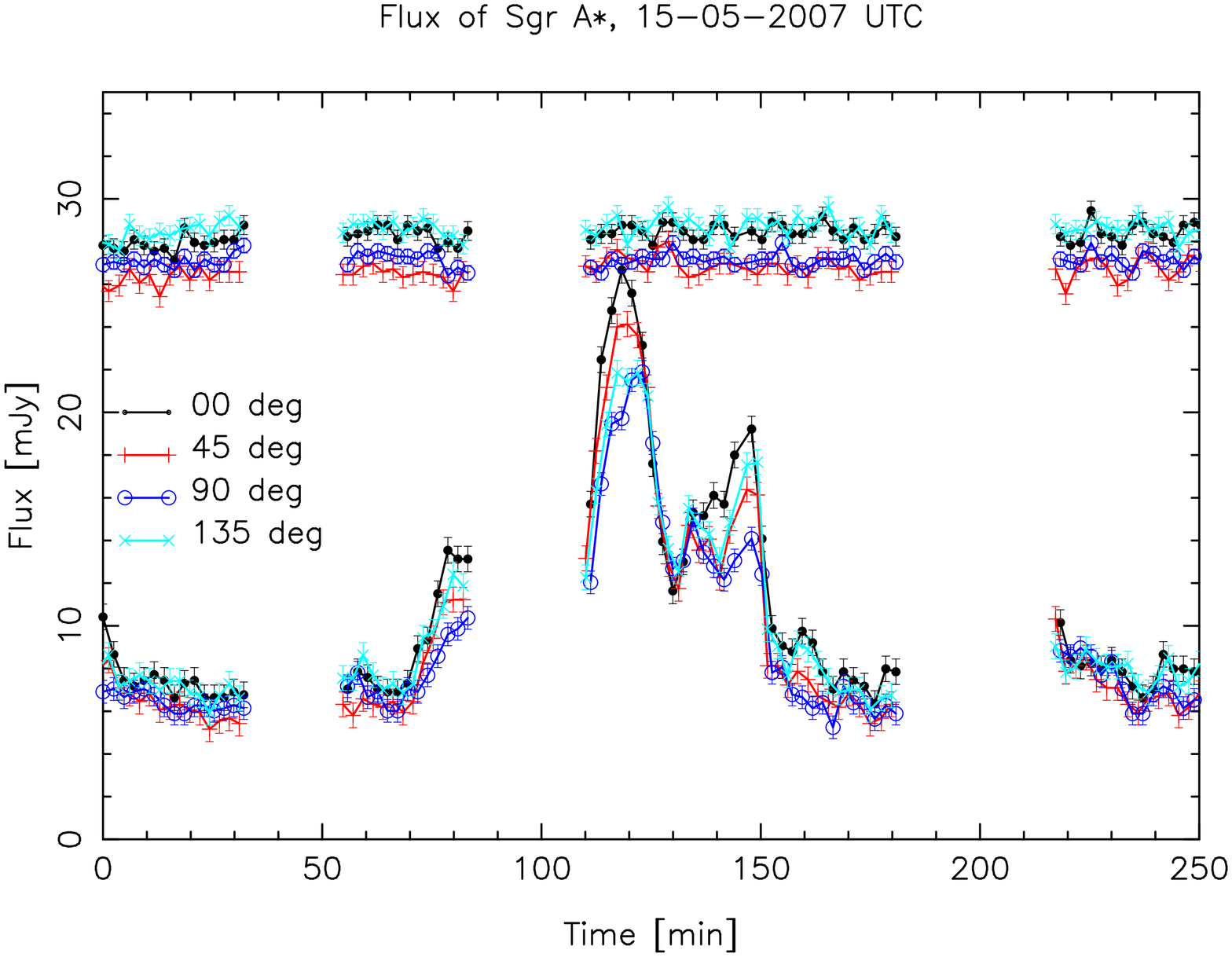}
\end{minipage}
\caption{Flux of the observed flares of Sgr~A* on   30 July 2005
(left) and 15 May 2007 (right) in different channels as a function
of time - each channel depicted in different color.  The light curve
of a constant star  is shown in the same plot and shifted by a few
mJy for a better view.} \label{moha-Fig0-2}
\end{center}
\end{figure}

\begin{figure}[t]
\hspace{4pc}
\begin{center}
\begin{minipage}{5cm}
\includegraphics[width=5.cm]{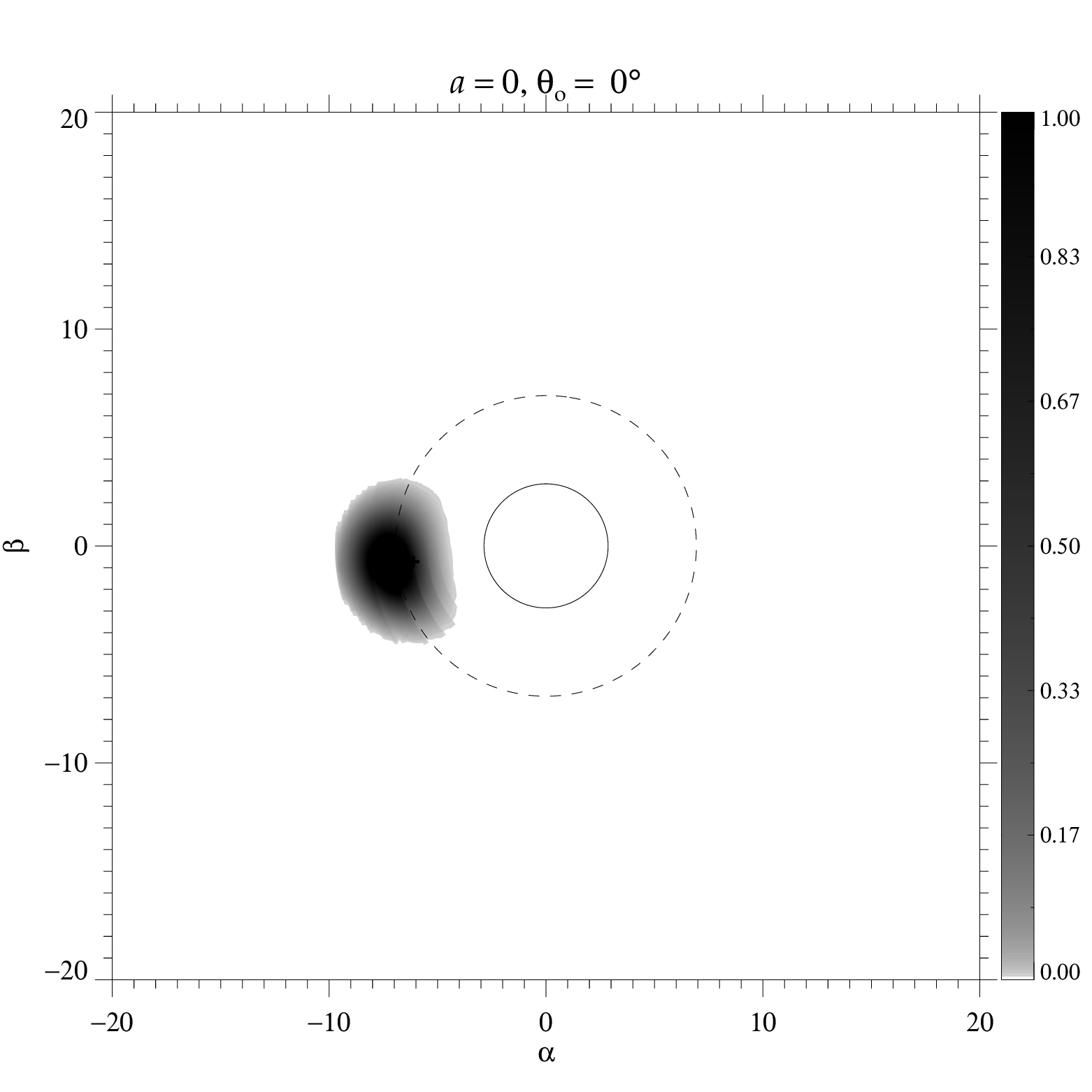}
\end{minipage}\hspace{2pc}%
\begin{minipage}{5cm}
\includegraphics[width=5.cm]{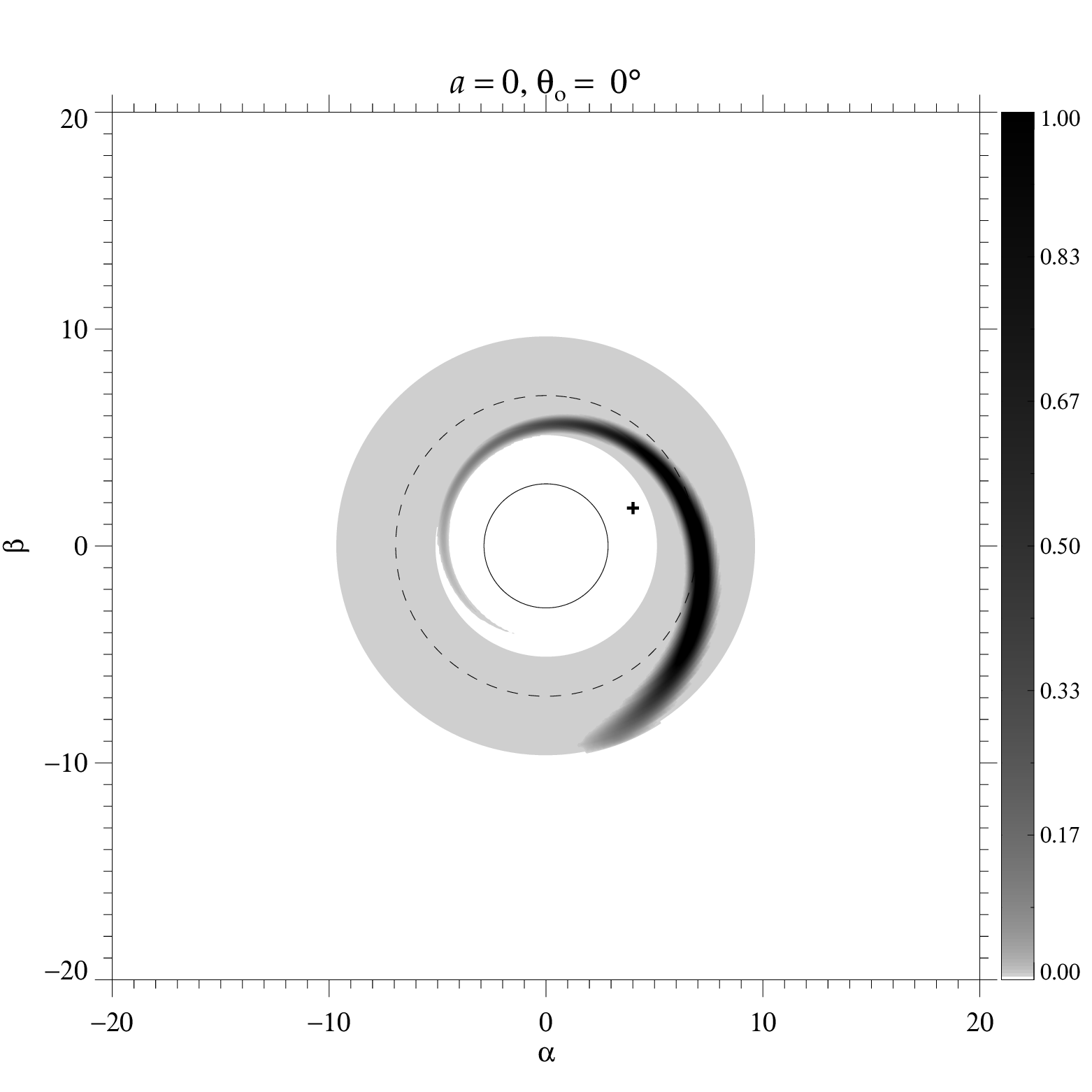}
\end{minipage}\hspace{0.5pc}
\begin{minipage}{5cm}
\includegraphics[width=5.cm]{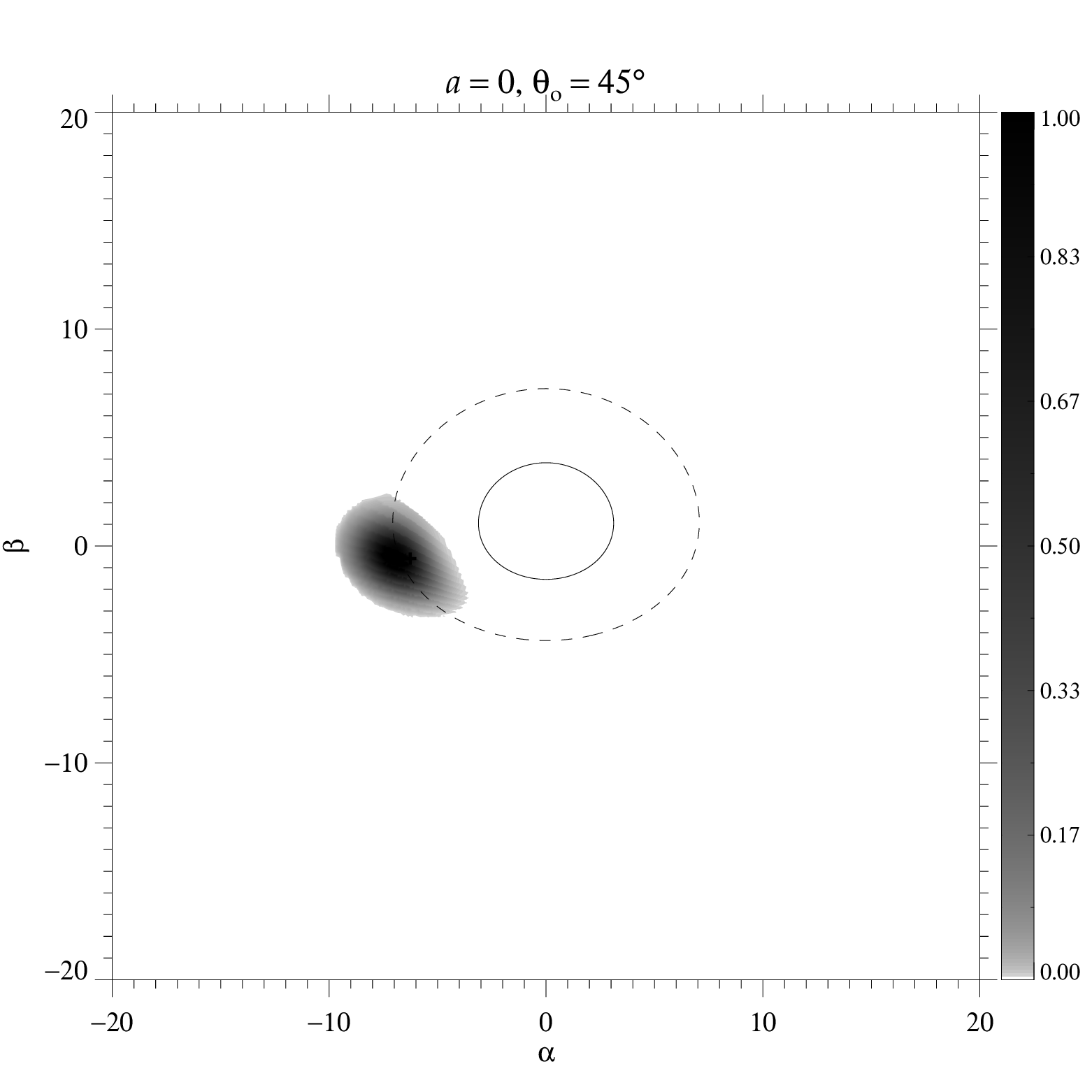}
\end{minipage}\hspace{2.pc}
\begin{minipage}{5cm}
\includegraphics[width=5.cm]{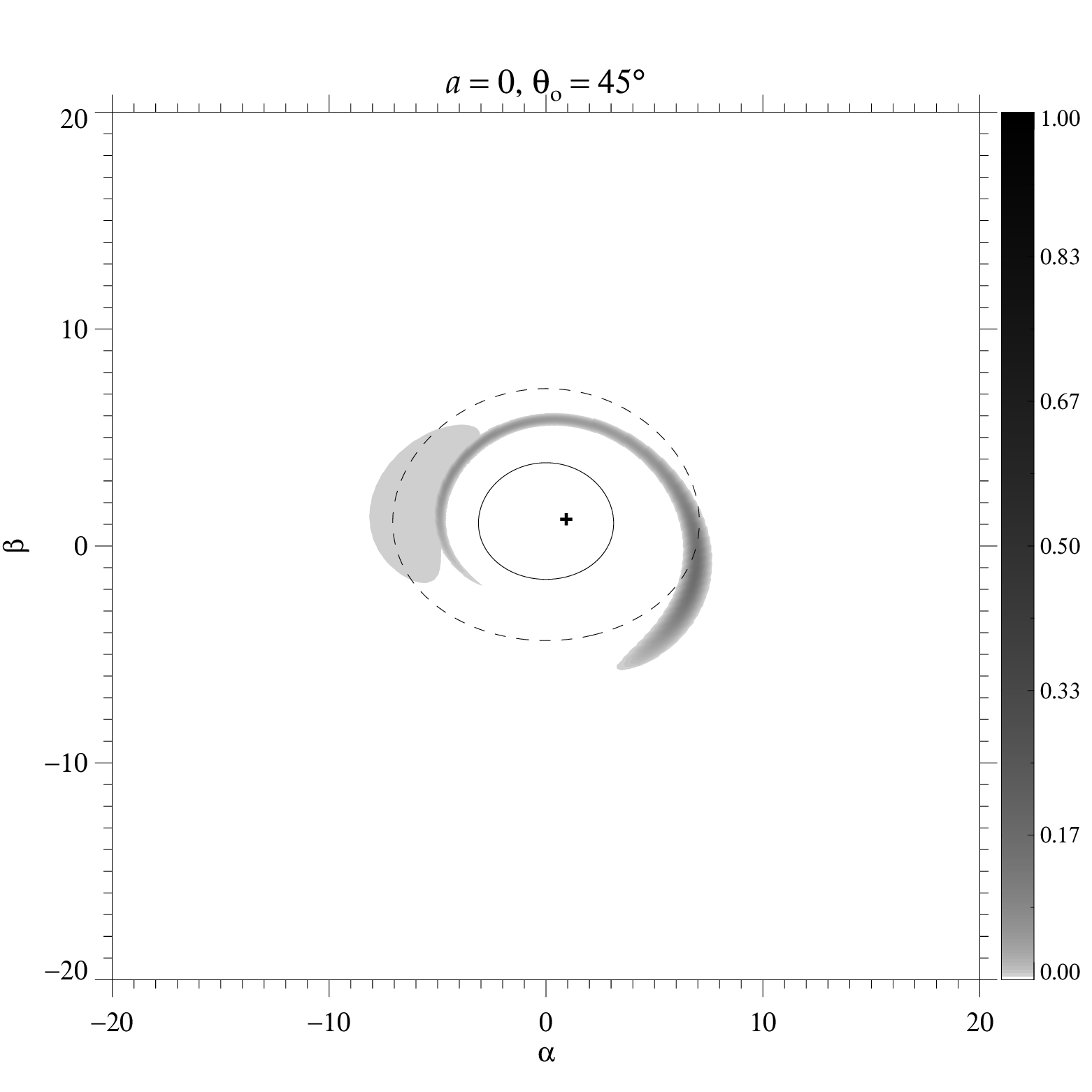}
\end{minipage}
\caption{The apparent images of an evolving spot in  Keplerian orbit around a Schwarzschild black hole,  plus a hot variable torus  projected onto the observer's image plane ($\alpha,\beta$) at inclination of $0^o$ (first row) and $45^o$ (second row) with respect to the common rotation axis.    Left column shows the beginnig of the event. Right column shows it after half an orbital time. The compact instability spreads radially via shock waves and produce a hot variable torus around the black hole. The surface brightness ratio of the torus to the spot is set to $\frac{1}{4}$, consistently with the results of Meyer et al. 2006a,b, Eckart et al. 2008. The spot cools down  due to synchrotron cooling and Keplerian shearing. Rotational shearing acts faster if the spot be located closer to the black hole (due to the larger differential velocity). Crosses  indicate the center of intesity of each image. The flux normalized to its maximum value for the whole orbit. Solid and dashed lines show the event horizon and marginally stable orbit, respectively. Labels are in $r_g$ units.} \label{moha-Fig0-3}
\end{center}
\end{figure}

The periods of enhanced radiation (so called "NIR flares",
normally around 100 minutes long) seem to be sometimes accompanied by quasi-periodic
oscillations (QPOs) (Fig.~\ref{moha-Fig0-2}; see also Genzel et
al. 2003, Eckart et al. 2004, 2006a,b, Meyer et al. 2006a,b). As we
will discuss later, in our modeling the main flare is caused by a
local event close to the Marginally Stable Orbit (MSO) which spreads
out via a shock wave and produces a temporary hot torus around the
black hole (Fig.~\ref{moha-Fig0-3}). The starting engine of the event
could be magnetic reconnections, stochastic acceleration of electrons
due to MHD waves or Magneto Rotational Instabilities (MRI) inside the
plasma.  The mentioned QPOs are mainly due to the presence of an
orbiting asymmetry in this torus which cause the flux modulations
according to general and special relativist effects.

Here we present two flares observed with the VLT in July 2005 and May
2007 and the results of our hot spot modeling which includes the
effects of rotational shearing.

\section{Observations and Data Reduction}
As part of a large observing campaign, Sgr~A* was observed in May
2007 and July 2005 using the VLT\footnote{Based on observations at
the Very Large Telescope (VLT) of the European Southern Observatory
(ESO) on Paranal in Chile; Programs: 073.B-0775 July 2004;
075.B-0093 July 2005; 079.B-0084 May 2007.}. The observations of
Sgr~A* were carried out in the NIR K$_S$-band (2.0-2.36$\mu$m) using
the  NIR camera CONICA and the adaptive optics (AO) module NAOS on
the European Southern Observatory's Very Large Telescope UT4 on
Paranal, Chile, during the nights between 29 and 30 July 2005 as
well as 14 and 15 May 2007. The infrared wavefront sensor of NAOS
was used to lock the AO loop on the NIR bright (K-band magnitude
$\sim$6.5) supergiant IRS~7, located about $5.6''$ north of Sgr~A*.
Therefore the AO  module was able to provide a stable correction with a
high Strehl ratio. In NACOS/CONICA (NACO) the
combination of a Wollaston prism with a half-wave retarder plate
allows the simultaneous measurement of two orthogonal directions of
the electric field vector.

The observed variable NIR emission of Sgr~A* has been found to be
usually highly polarized and to consist of a contribution of a non- or
weakly polarized main flare with highly polarized sub-flares, showing
a possible QPO of 17$\pm$3 minutes consistent with previous
observations. Significant positive flux density excess on time scales
shorter (according to all previous NACO observations typically of the
order of the QPO time scale) than the overall flare duration
(typically 100 minutes or more) are called sub-flares. In
Fig.~\ref{moha-Fig0-2} we show the flux density per polarization
channel as a function of time.  The 3 gaps in the data are due to sky
observations.  For more details on the observations and data reduction
see Eckart et al.  2006b, 2008.

\section{Modeling and Discussion}

As mentioned above we interpret QPOs according to the existence of an
orbiting asymmetry in the flaring region of the accretion disk close
to the MSO of the black hole. The fact that the frequency of these
QPOs is comparable to the MSO's dynamical time for  a black hole
with the mass of Sgr~A*, gives more strength to the assumption of an
orbiting bright region (so called "spots"). In our modeling we
simulate the relativistic effects by using the KY code (Dovciak et
al. 2004). It takes into account the relativistic effects
(relativistic beaming, redshifts and blue-shifts, lensing, time
delays, change of the emission angle and change of polarization angle)
using a slim disk approximation. This allows us to fit the model
parameters to the actual data. The procedure was first demonstrated by
Meyer et al. (2006a,b; the authors also discuss in detail the
differences with other modeling efforts, e.g. Broderick \& Loeb
2006a,b). The magnetic field configuration is such that the resulting
projected E-vector is always perpendicular to the equatorial plane
(see also Shakura \& Sunyaev 1973). As a second configuration we have
allowed for a global toroidal magnetic field.

By using a simple approach which considers only the rise and fall
times of the observed flux, one can put an upper limit on the
characteristic size of the emitting region. For a Shakura-Sunyaev-type
accretion disk and in case that self gravity effects can be neglected
(very close to the black hole), we can use the standard one-zone
approximation of the vertical hydrostatic equilibrium. This
approximation relates the azimuthal velocity, the sound velocity, the
thickness of the disk, and the radius to each other (see the
discussion by Meyer et al. 2006). In order to assign an upper limit to
the size of the emitting region $\Delta r$, we use the minimum of a
stable dynamical time related to the maximally rotating black hole
($a=1$), $t_d\simeq 4 $ \ min, and typical observed rise time of the
flux, $t_{r} \simeq 40$ min. For a geometrically thin disk
($\frac{h}{r}\sim 10^{-1}$) this leads to $\Delta r\leq r_{mso}$.

Evidence from MHD simulations indicates that the dynamical transition
from rotational support to free fall which occurs in the vicinity of
the MSO could be the source of magnetic or thermal instabilities (Krolik
\& Hawley 2002). Together with the assumption that the radiation edge
of the accretion disk is located very close to the marginally stable
orbit (Krolik \& Hawley 2002), one could put a constraint on the spin
of the black hole. If the observed 20 min NIR modulations are related
to the orbiting spots, the radius of their orbit must be much smaller
than the marginally stable orbit of a Schwarzschild black hole. The
above considerations imply that the dimensionless spin parameter
reaches a minimum value of 0.5 (see also Eckart et al. 2004, Meyer et
al. 2006a,b).

\begin{figure}[t]
\hspace{1pc}
\begin{center}
\begin{minipage}{5cm}
\includegraphics[width=5cm,angle=270]{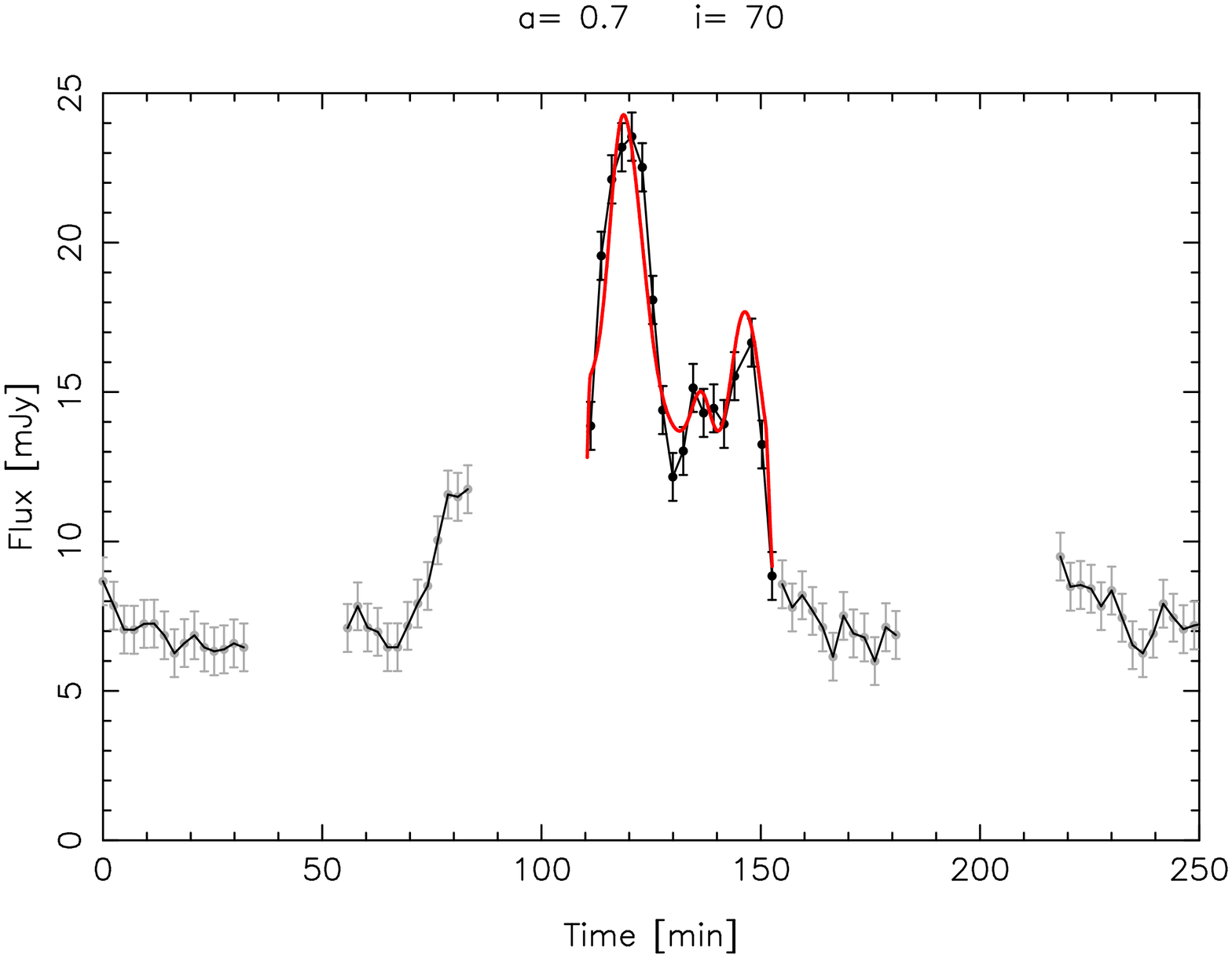}
\end{minipage}\hspace{4pc}
\begin{minipage}{5cm}
\includegraphics[width=5cm,angle=270]{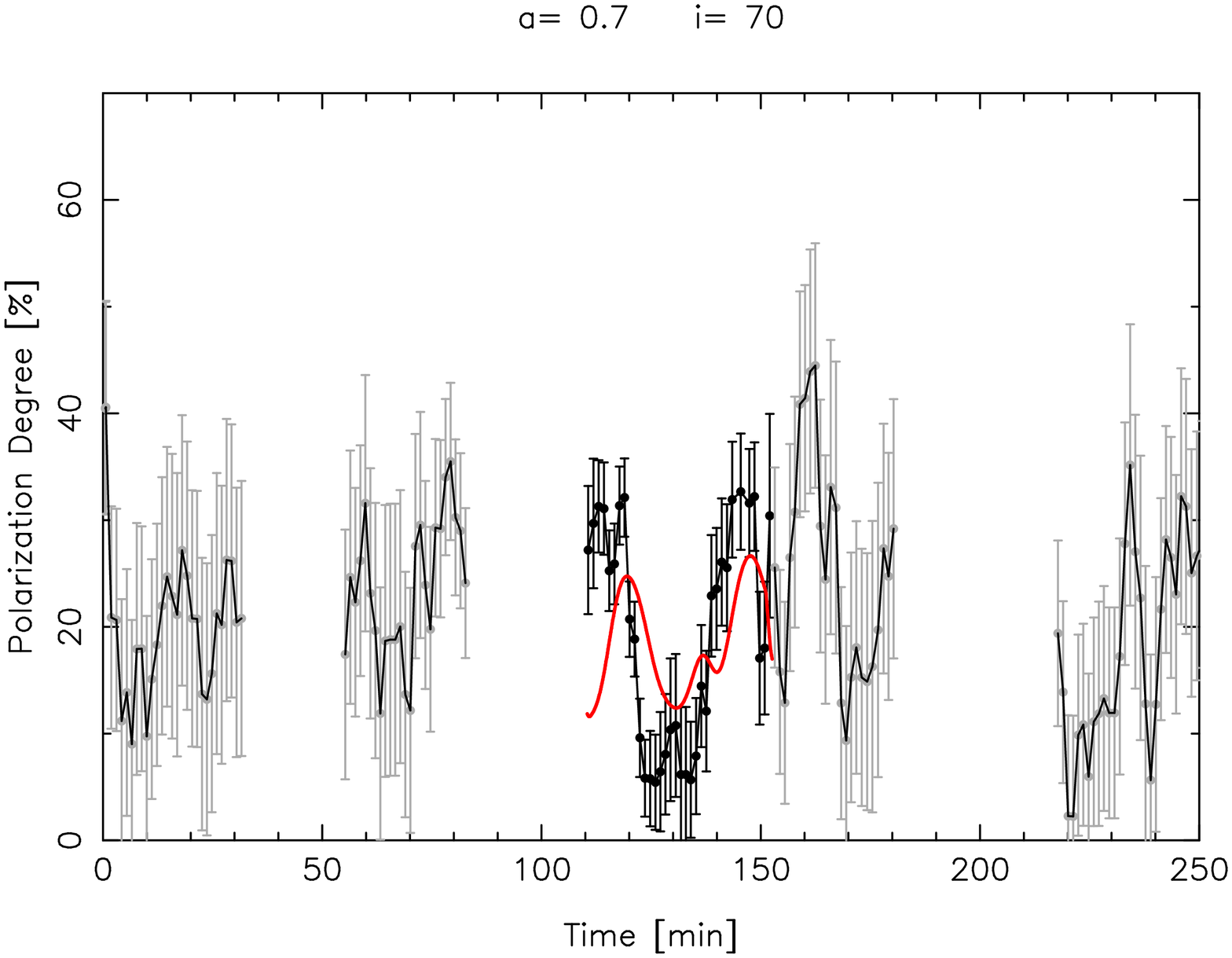}
\end{minipage}
\caption{The best fit to the observed total flux density (left) and degree of linear polarization (right), achieved by the  spot model including the
rotational shearing for  the 15 May 2007 event. } \label{moha-Fig0-4}
\end{center}
\end{figure}

Since we focus only on modeling of the NIR flares, it is sufficient to
pick up the simple power-law model for the non-thermal electrons
energy distribution: $N(\gamma) = N_0\gamma^{-p}$, with a sharp
cut-off energy $\gamma_c$. This leads to a simple formula for the
synchrotron emission: $I_\nu \ \propto \ n
\ {B_{\perp}}^{(\frac{p+1}{2})}\nu^{-(\frac{p-1}{2})}$, where $N_0$,
$n$, $p$ and $B_{\perp}$ are the normalization constant, the number
density of the electrons, the spectral energy index and perpendicular
component of the magnetic field projected onto the plane of an observer
co-moving with the plasma, respectively. $I_\nu$ is the first Stokes
parameter. Since it is proved observationally that there exists a
turn-over frequency in the sub-millimeter to NIR range, by using the
turn-over ferquency formula one could assign an upper limit on
$\gamma_c^2B_{\perp}$. Here we used $\gamma_c=100, B=60$G which gives
the best fit to the existing NIR/X-ray flare models (Liu et al. 2006,
Eckart et al. 2008). The spectral energy index is fixed to $p=2.2$,
according to recent observations by Hornstein et al. 2006.

\begin{figure}[t]
\hspace{4pc}
\begin{center}
\begin{minipage}{5cm}
\includegraphics[width=5.cm]{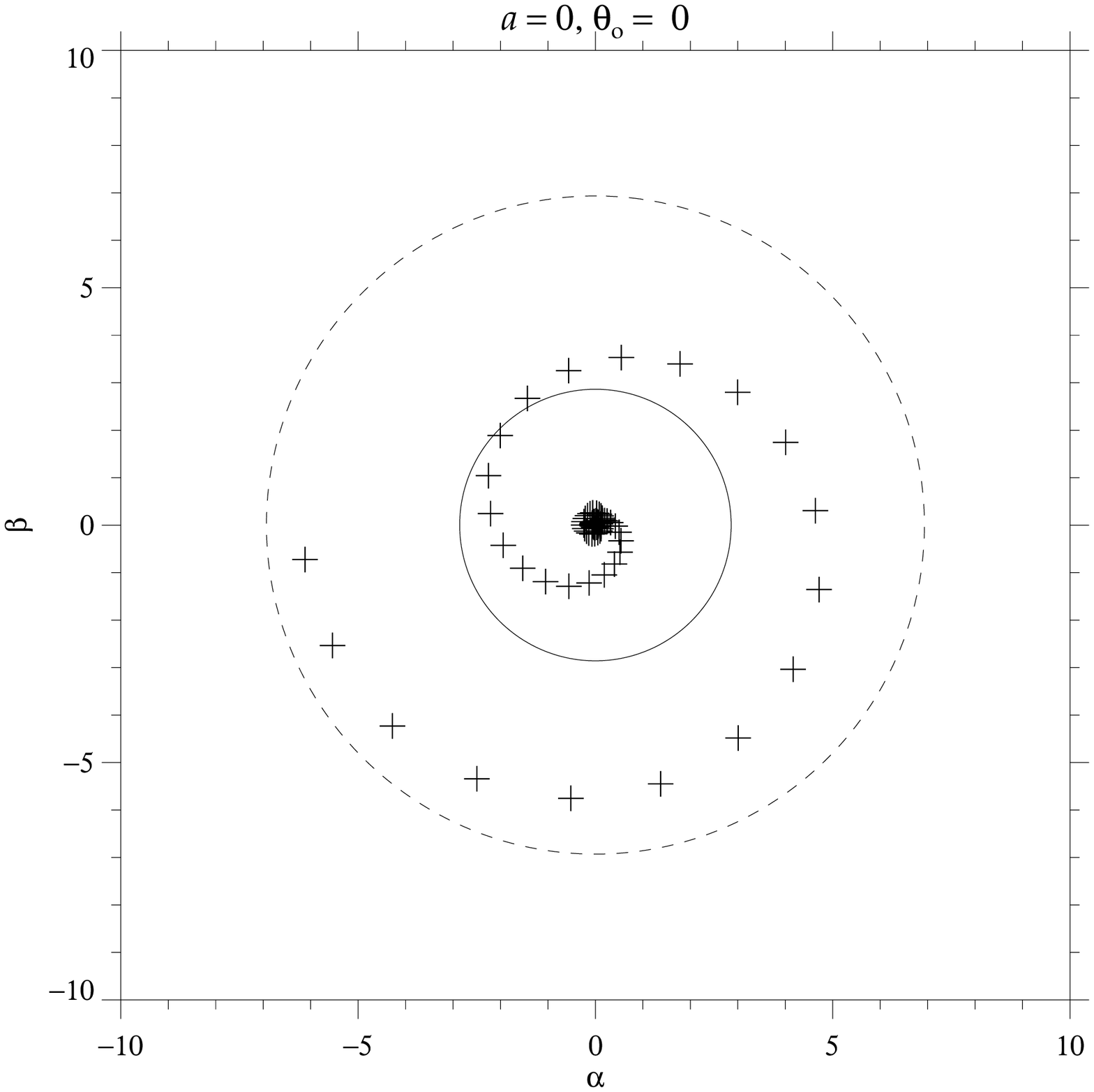}
\end{minipage}\hspace{2pc}%
\begin{minipage}{5cm}
\includegraphics[width=5.cm]{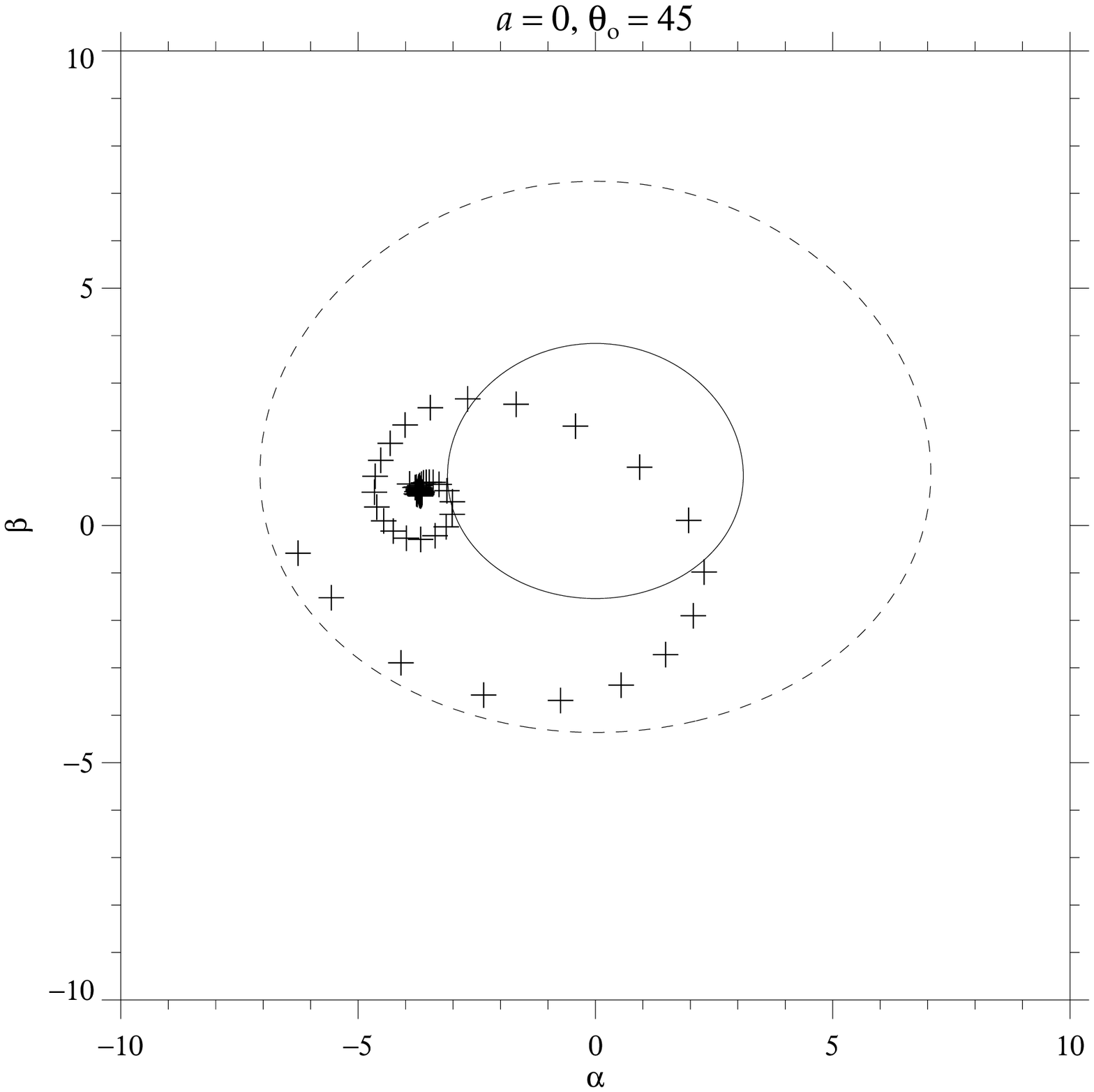}
\end{minipage}\hspace{0.5pc}
\begin{minipage}{5cm}
\includegraphics[width=5.cm]{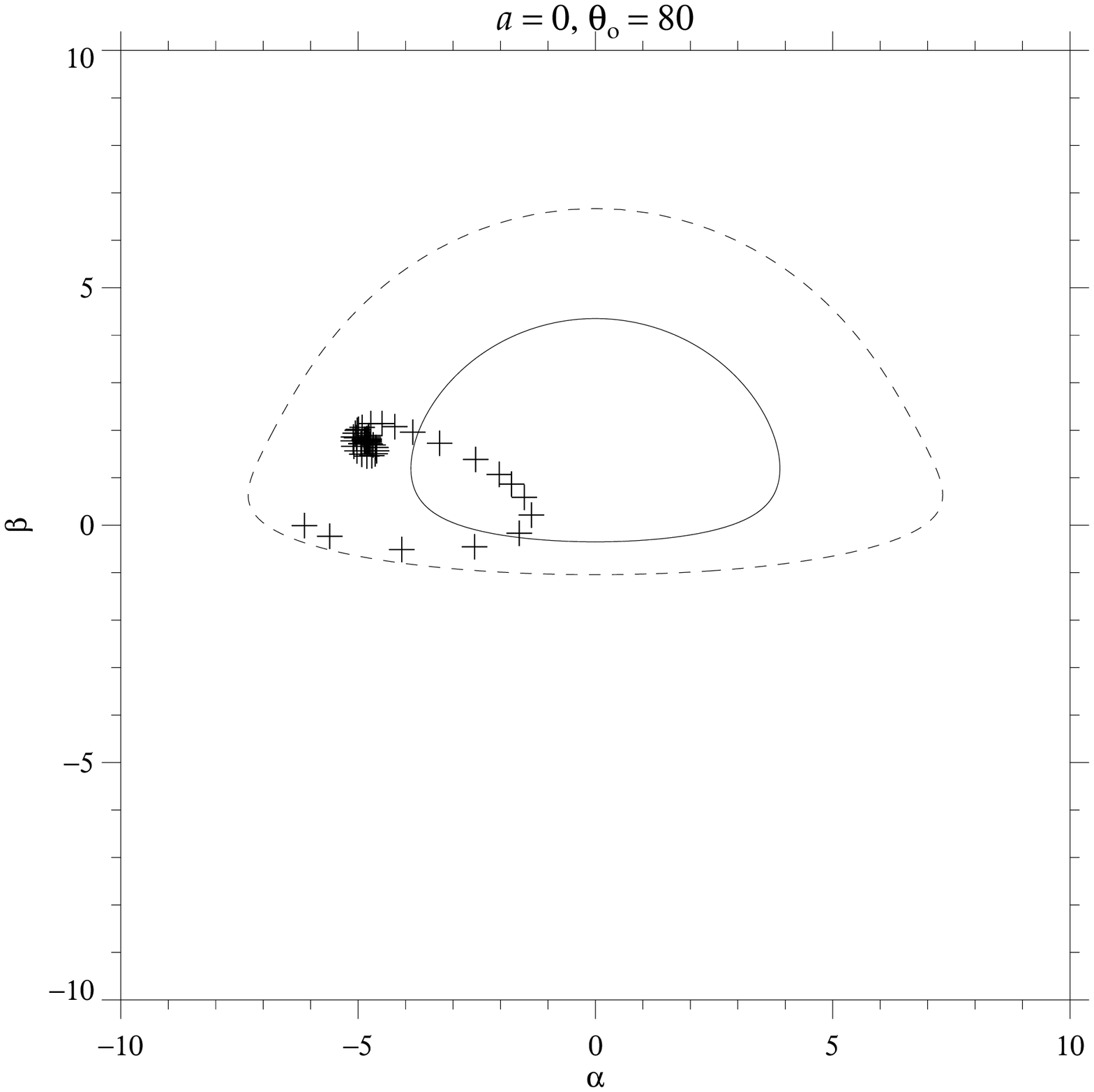}
\end{minipage}\hspace{2.pc}
\begin{minipage}{5cm}
\includegraphics[width=5.cm]{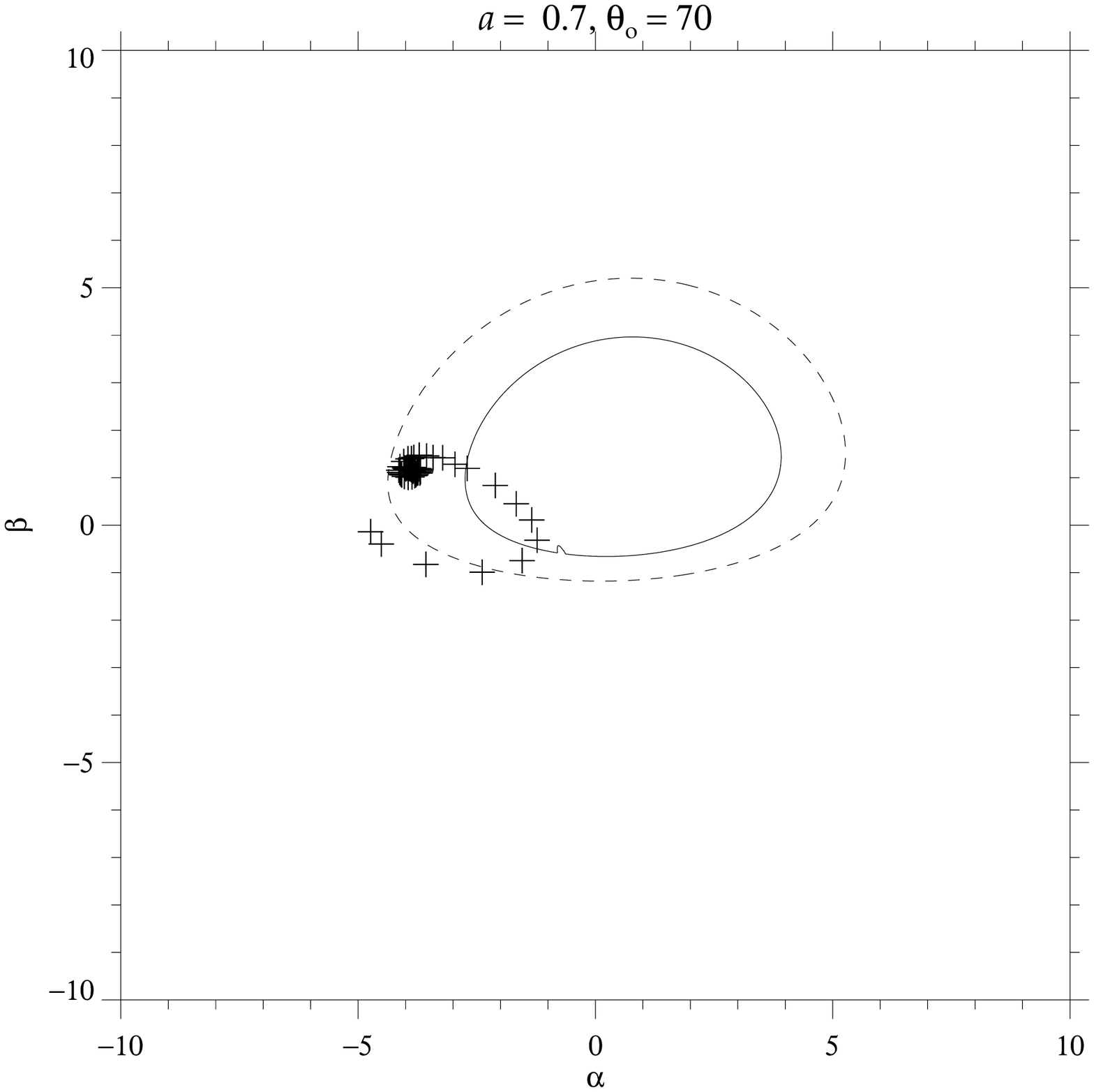}
\end{minipage}
\caption{The centroid motion of the NIR images projected onto the observer's image plane
($\alpha,\beta$) as seen for inclination $0, 45$ and $80$ degrees for a Schwarzschild black hole. Each image shows four revolutions of a spot in the presence of Keplerian shearing. Time lag between each cross is $\frac{1}{20}$ of one orbital time.   Note that in higher inclinations the boosted photons from the torus cause an offset from the center to the left. Right bottom: the centroid track for the best fit parameters to the 15 May 2007 event. Labels are in $r_g$ units. The solid and dashed lines show the event horizon and the MSO, respectively. } \label{moha-Fig0-5}
\end{center}
\end{figure}

 Fig.~\ref{moha-Fig0-4}  shows
the least $\chi^2_{reduced}$  fit to the NIR flare observed on 15th of May 2007 via a simple approach (see Eckart et al. 2008). 
 Here we fitted the flux and intensity of polarization for
the spin parameter $a$, the inclination angle $i$, the over-brightness of the spot
according to the disk, the polarization degrees  of the torus
(restricted between 0\%-5\%) and the spot (restricted between
0\%-70\%), as well as the initial phase of the spot on the orbit.
The inclination $i$ is defined such that the temporary accretion
disk is seen edge on for an inclination of $i$=90$^o$. The upper limit for the spot polarization reflects
the maximum value that could be produced by synchrotron radiation. Fig.~\ref{moha-Fig0-4} clearly shows that this model can
produce the substructures in flux and also in polarization degree.
The fit prefers a high inclination angle  and spin parameter
($i=70^o$ \& $a=0.7$; $\chi_{reduced }^2=3.55$), which is also in good
agreement with previous works by Eckart et al. (2006b) and
Meyer et al. (2006a,b). 

As discussed above, the May 2007 data possibly represent the first
direct observational evidence that spots may only be stable for about
one orbital time scale. An efficient creation of spots could be
provided through magneto-rotational instabilities that are shown to be
present in Keplerian rotating accretion disks even in the presence of
a dominating toroidal magnetic field (Hawley \& Balbus 1991). Such
shear-flow instabilities are a fast mechanism to generate a turbulent
flow in a Keplerian disk (e.g.  three-dimensional simulations by Arlt
\& R\"udiger 2001).

The next generation VLTI instrument GRAVITY (Gillessen et al. 2006b)
would make it possible to track the NIR centroid path of Sgr~A* with
$\approx 10 \mu$as $\equiv 1 r_g$ precision. Fig.~\ref{moha-Fig0-5}
shows the expected centroid tracks of an evolving spot for different
inclinations of the accretion flow with respect to the line-of-sight
for the case of a spinless black hole and for the best fit
parameters of the 15 May 2007 event. The rotational shearing and the
Doppler boosted photons from the torus make the centroid path more
compact in comparison to background subtracted images of an orbiting
spot with a solid shape (see Broderick \& Loeb 2006a,b). This may present
a complication in detecting the plasma structure close to the event
horizon of Sgr~A*.

\section*{Acknowledgments} Part of this work was supported by the
German \emph{Deut\-sche For\-schungs\-ge\-mein\-schaft, DFG\/} via
grant SFB 494. M. Zamaninasab and D. Kunneriath
 are members of the International
Max Planck Research School (IMPRS) for  Astronomy and Astrophysics
at the MPIfR and the Universities of Bonn and Cologne.

\end{document}